\documentclass{amsart}
\usepackage{amsmath, amssymb}

\newcommand{\llq}{\lq\lq} 

\newcommand{\N}{\mathbb N} 
\newcommand{\Z}{\mathbb Z} 
\newcommand{\R}{\mathbb R} 
\newcommand{\C}{\mathbb C} 

\providecommand{\vect}[1]{\mathbf{#1}} 
\providecommand{\matr}[1]{\mathrm{#1}} 

\newtheorem{thm}{Theorem}[section]
\newtheorem{defn}[thm]{Definition}
\newtheorem{prop}[thm]{Proposition}
\newtheorem{lemma}[thm]{Lemma}
\newtheorem{cor}[thm]{Corollary}
\newtheorem{rmrk}[thm]{Remark}

\numberwithin{equation}{section}
\newcommand{\prf}{\textbf{\underline{Proof}:} }

\begin{document}

\title{On the Stokes matrices of the $tt^*$-Toda equation}
\author{Stefan Horocholyn}

\begin{abstract}
We derive a formula for the signature of the symmetrized Stokes matrix $\mathcal{S}+\mathcal{S}^\matr{T}$ for the $tt^*$-Toda equation.  As a corollary, we verify a conjecture of Cecotti and Vafa regarding when $\mathcal{S}+\mathcal{S}^\matr{T}$ is positive definite, reminiscent of a formula of Beukers and Heckmann for the generalized hypergeometric equation.  The condition $\mathcal{S}+\mathcal{S}^\matr{T} > 0$ is prominent in the work of Cecotti and Vafa on the $tt^*$ equation; we show that the Stokes matrices $\mathcal{S}$ satisfying this condition are parameterized by the points of an open convex polytope.
\end{abstract}

\maketitle

\section{Introduction}

The classical Stokes phenomenon for meromorphic ODE has begun to play an important role in geometry, notably in singularity theory, Frobenius manifolds and mirror symmetry.  For a (real) Stokes matrix $\mathcal{S}$, the symmetrized matrix $\mathcal{S}+\mathcal{S}^\matr{T}$ arises in the context of Frobenius manifolds and the $tt^*$ equation (e.g. \cite{Du96, Du99, HeSa11, HeSe07}).

The $tt^*$ equation is a system of nonlinear PDEs which appeared in the work of Cecotti and Vafa \cite{CV1, CV2, CV3} on the classification of supersymmetric field theories in physics. It is a special case of the harmonic map equation in differential geometry for maps from a surface to a noncompact symmetric space \cite{GL1}.  Dubrovin \cite{Du93} showed that it admits an isomonodromic deformation interpretation, as well as a zero-curvature formulation.  This leads to a Riemann-Hilbert correspondence between (local) solutions and monodromy data of a meromorphic ODE. Clarifying this correspondence is a subject of current research activity relating several fields of mathematics, including Hodge theory and algebraic geometry.

There are very few examples where solutions can be found.  A special case of the $tt^*$ equation, introduced by Cecotti and Vafa, and studied mathematically by Guest-Its-Lin \cite{GIL1, GIL2} and Mochizuki \cite{MoXX}, are the $tt^*$-Toda equation.  This is, essentially, the well-known Toda field equation (2-dimensional periodic Toda lattice), although even in this case the existence of the solutions predicted by Cecotti and Vafa was proved only recently (in the aforementioned references).

This article was motivated by the conjectures of Cecotti and Vafa regarding the symmetrized Stokes matrix $\mathcal{S}+\mathcal{S}^\matr{T}$, in the case of the $tt^*$-Toda equation.  We shall give a necessary and sufficient condition for $\mathcal{S}+\mathcal{S}^\matr{T}$ to be positive definite, and a simple formula for the signature of $\mathcal{S}+\mathcal{S}^\matr{T}$, in general, which is reminiscent of a formula of Beukers and Heckmann for the generalized hypergeometric equation \cite{BH}.

Let us now state the $tt^*$-Toda equation and explain the relevant Stokes matrix.  The equations are:
\begin{equation}\label{tt*-TodaEqtn}
2(w_i)_{z\overline{z}} = -e^{2(w_{i+1} - w_i)} + e^{2(w_i - w_{i-1})} \ , \quad w_i: \C^* \rightarrow \R
\end{equation}
subject to two further conditions:
\begin{enumerate}
\item the $\llq$anti-symmetry" condition: $w_i + w_{n-i} = 0$; and
\item the radial condition: $w_i = w_i(|z|)$.
\end{enumerate}
We assume that $w_i = w_{i+n+1}$ for all $i \in \Z$. In what follows, we write $n+1=2m$ or $n+1=2m+1$ (i.e. $m := \lfloor \tfrac{n+1}{2} \rfloor$).

This system is the compatibility condition for the linear system:
\[ \left\{\begin{array}{ccl}
       \Psi_z  & = & (\matr{w}_z + \tfrac{1}{\lambda} \matr{W})\Psi \ , \\
\Psi_{\overline{z}} & = & (-\matr{w}_{\overline{z}} + \lambda \matr{W^T})\Psi \ , \end{array}\right. \]
where:
\[ \matr{w} = \matr{diag} \, (w_0, \dots, w_n) \ , \quad \matr{W} = \left[ \begin{array}{c|c|c|c}
\vphantom{ (w_0)_{(w_0)}^{(w_0)} } & \! e^{w_1 \! - \! w_0} \! &        &                                          \\ \hline
                                   &                           & \ddots &                                          \\ \hline
\vphantom{\ddots}                  &                           &        & e^{w_n \! - \! w_{n \! - \! 1}} \! \! \! \\ \hline
\vphantom{ (w_0)_{(w_0)}^{(w_0)} } \! \! e^{w_0 \! - \! w_n} \! \! & &  & \! 
\end{array} \right] \ . \]
If we write $x = |z|$, then the radial version of (\ref{tt*-TodaEqtn}) is the compatibility condition for a linear system, which may then be transformed to (see Equation 1.4 of \cite{GIL2}):
\begin{equation}\label{RadialCompatEqtn}
\begin{cases} \Psi_\zeta & = \ \left( -\tfrac{1}{\zeta^2} \matr{W} - \tfrac{1}{\zeta} x\matr{w}_x + x^2\matr{W^T} \right) \Psi \ , \\
              \Psi_x     & = \ \left( \tfrac{1}{x\zeta} \matr{W} + x\zeta \matr{W^T} \right) \Psi \ , \end{cases}
\end{equation}
where $\zeta = \tfrac{\lambda}{z}$.

The equation for $\Psi_\zeta$ in (\ref{RadialCompatEqtn}) is a meromorphic linear ODE in the complex variable $\zeta$, with poles of order two at both $\zeta=0$ and $\zeta=\infty$.  The Stokes matrices at these two poles are equivalent, so we shall only consider the Stokes matrix at $\zeta = \infty$, and denote it by $\mathcal{S}$.  By the general theory of isomonodromic deformations (e.g. \cite{FIKN}), Stokes matrices $\mathcal{S}$ correspond to local solutions near $0$ (i.e. defined on intervals of the form $(0,\varepsilon)$) of the $tt^*$-Toda equation.  Further details and explanation may be found in \cite{GIL1, GIL2}, where $\mathcal{S}$ is computed in terms of the asymptotic behaviour of the functions $w_i$.

It was conjectured by Cecotti and Vafa that the condition $\mathcal{S}+\mathcal{S}^\matr{T} > 0$ implies that the corresponding local solution of the $tt^*$-Toda equation is globally defined on $\C^*$ (i.e. such that $\varepsilon=\infty$).  This was confirmed in \cite{GIL1, GIL2, GL1, MoXX}, and in Theorem 5.6 of \cite{GIL2}, a stronger result (also suggested by Cecotti and Vafa) was shown: a necessary and sufficient condition for the local solution of the $tt^*$-Toda equation to be globally defined on $\C^*$ is that the eigenvalues of the monodromy $\mathcal{S}(\mathcal{S}^{-1})^\matr{T}$ are of unit length.

It is, therefore, of interest to describe the set of such Stokes matrices explicitly, and this is our first main result. For such Stokes matrices, we prove the following explicit characterization of the signature of $\mathcal{S}+\mathcal{S}^\matr{T}$, showing that they form an open convex polytope described by simple equations, and we expect this result to be of use in future investigations of the $tt^*$-Toda equation:

\textbf{\underline{Theorem}:} \emph{$\mathcal{S}+\mathcal{S}^\matr{T}$ has the same signature as the diagonal matrix:}
\[ \matr{diag} \, ( (-1)^{n+1} p(\pi_0 ), \dots, (-1)^{n+1} p(\pi_{n}) ) \ . \]
\emph{Here, $\pi_k$ are the $n+1$ roots of $x^{n+1} - (-1)^{n+1}$, and the real polynomial $p(x)$ is the characteristic polynomial of a certain matrix $\mathcal{R}$ satisfying $(-1)^n\mathcal{R}^{n+1} = \mathcal{SS}^\matr{-T}$.}

\textbf{\underline{Corollary}:} \emph{$\mathcal{S}+\mathcal{S}^\matr{T} > 0$ \emph{iff} $(-1)^{n+1} p(\pi_k ) > 0$ for all $k$, and the set of such Stokes matrices is in 1-1 correspondence with an open, convex polytope of $\R^m$.}

Our second main result is a formula for the sign of $p(\pi_k)$ when the eigenvalues of $\mathcal{S}(\mathcal{S}^{-1})^{\matr{T}}$ are unimodular.  We refer the reader to Corollary \ref{TheFormula} for the precise statement of this result.  This characterizes the signature of $\mathcal{S}+\mathcal{S}^\matr{T}$ in terms of the configurations of the eigenvalues of $\mathcal{R}$ with respect to the roots $\pi_k$ on the unit circle.

These results are given in Section 2 for a conveniently defined, $\llq$idealized Stokes matrix" $\matr{S}$.  In Section 3, we explain the precise relation between this $\llq$idealized Stokes matrix", and the $\llq$actual" Stokes matrices of \cite{GIL1, GIL2}.

Notational remark: In this paper, $\N$ shall denote the natural numbers, $\Z$ the integers, $\Z_{\geq 0}$ the non-negative integers, $\R$ the real numbers, $\C$ the complex numbers, and $\C^* = \C \setminus \{ 0 \}$ the complex plane punctured at the origin. For a matrix $\matr{A}$, its transpose is denoted $\matr{A^T}$, and $\matr{A^{-T}}$ will denote the inverse of $\matr{A^T}$. \\


\section{Main Results}

Let $\matr{R} \in \matr{SL}_n\R$ be given by:
\begin{equation}\label{RMatrix}
\matr{R} := \begin{bmatrix} -p_{n-1}  &  \cdot &     \cdots     &  \cdot \\
                             \vdots   & \vdots & \matr{I}_{n-1} & \vdots \\
                              -p_1    &  \cdot &     \cdots     &  \cdot \\
                            -\epsilon &    0   &     \cdots     &    0    \end{bmatrix} \ ,
\end{equation}
where $\epsilon := (-1)^n$, and where the characteristic polynomial $p(x)$ of $\matr{R}$,
\[ p(x) = x^n + \sum_{k=1}^{n-1} p_k \, x^k + \epsilon \ , \]
is \emph{signed-palindromic}:
\begin{equation}\label{Palindromic}
p(x) = (-x)^n p\left( \tfrac{1}{x} \right) \quad \Leftrightarrow \quad p_{n-k} = \epsilon p_k \quad \forall \ 1 \leq k \leq n-1 \ .
\end{equation}
In addition, let $\matr{S}$ be the upper-triangular Toeplitz matrix:
\begin{equation}\label{StokesMatrix}
\matr{S} := \begin{bmatrix}    1   & \epsilon p_1  & \cdots & \epsilon p_{n-2} & \epsilon p_{n-1} \\
                               0   &    1   & \ddots & \ddots  & \epsilon p_{n-2} \\
                            \vdots & \ddots & \ddots & \ddots  & \vdots  \\
                            \vdots & \ddots & \ddots &    1    & \epsilon p_1    \\
                               0   & \cdots & \cdots &    0    &   1     \end{bmatrix} \ .
\end{equation}

\begin{prop}\label{NthRoot}
$-\epsilon \matr{R}^n = \matr{SS^{-T}}$.
\end{prop}

\prf We show that $-\epsilon \matr{R}^n$ has an upper-lower-triangular decomposition:
\[ -\epsilon\matr{R}^n = \matr{UL} \ , \]
where $\matr{U}$ and $\matr{L}$ have only 1s on the diagonal, and in doing so, will show that $\matr{U} = \matr{S}$ and $\matr{L}^{-1} = \matr{S^T}$.  We first apply $-\epsilon\matr{R}^n$ and $\matr{UL}$ to the flag:
\[ \mathcal{F} \ : \ \vect{0} \subset \mathcal{F}_0 \subset \mathcal{F}_1 \subset \cdots \subset \mathcal{F}_{n-1} \cong \R^n \ , \quad \mathcal{F}_k := \, < \vect{e}_n, \vect{e}_{n-1}, \dots, \vect{e}_{n-k+1}, \vect{e}_{n-k}>_\R \ , \]
where $\vect{e}_k$ is the $k^\matr{th}$ canonical unit vector.  By inspection, $\matr{L}$ fixes $\mathcal{F}$, so it suffices to compare respective applications of $-\epsilon\matr{R}^n$ and $\matr{U}$ to $\mathcal{F}$.

The proof is facilitated by two observations: the first is that $\matr{R}\vect{e}_k = \vect{e}_{k-1}$ for all $2 \leq k \leq n$, which implies that $\vect{e}_k = \matr{R}^{n-k} \vect{e}_n$ for all $1 \leq k \leq n$.  The second observation is that, by Cayley-Hamilton:
\[ \matr{R}^n + \sum_{k=1}^{n-1} p_k \matr{R}^k + \epsilon\matr{I}_n = 0 \ , \]
which, taken together with the first observation, yields:
\[ \matr{R}^n\vect{e}_n = -p_{n-1}\vect{e}_1 - \cdots - p_1\vect{e}_{n-1} - \epsilon\vect{e}_n \ . \]

The proof proceeds as follows: applying $-\epsilon\matr{R}^n$ and $\matr{U}$ to $\vect{e}_n \in \mathcal{F}_0$ yields:
\[ \epsilon p_{n-1}\vect{e}_1 + \cdots + \epsilon p_1\vect{e}_{n-1} + \vect{e}_n = -\epsilon\matr{R}^n\vect{e}_n = \matr{U}\vect{e}_n = u_{1,n}\vect{e}_1 + \cdots + u_{n-1,n}\vect{e}_{n-1} + \vect{e}_n \ , \]
which determines the last column of $\matr{U}$. Next, to determine the second-last column, we observe that, on the one hand:
\[ \matr{R}^n\mathcal{F}_1 = \matr{R}^n\mathcal{F}_0 + \matr{R}^n<\vect{e}_{n-1}>_\R \ , \]
and on the other hand, $<\matr{R}\vect{e}_1>_\R = \matr{R}^n\mathcal{F}_0$ by the above observations.  Hence, it follows that:
\begin{align*}
-\epsilon\matr{R}^n\vect{e}_{n-1} 
 & = \epsilon\matr{R}\big( p_{n-1}\vect{e}_1 + p_{n-2}\vect{e}_2 + \cdots + p_1\vect{e}_{n-1} + \epsilon\vect{e}_n \big) \\
 & \equiv \epsilon\matr{R}\big( p_{n-2}\vect{e}_2 + \cdots + p_1\vect{e}_{n-1} + \epsilon\vect{e}_n \big) \mod \matr{R}^n\mathcal{F}_0 \\
 & = \epsilon p_{n-2}\vect{e}_1 + \cdots + \epsilon p_1\vect{e}_{n-2} + \vect{e}_{n-1} \ .
\end{align*}
By comparing this with the application of $\matr{U}$ to $\vect{e}_{n-1} \in \mathcal{F}_1$:
\[ \matr{U}\vect{e}_{n-1} = u_{1,n-1}\vect{e}_1 + \cdots + u_{n-2,n-1}\vect{e}_{n-2} + \vect{e}_{n-1} \ , \]
we see that we have determined the second-last column of $\matr{U}$. Continuing this process, at the $k^\matr{th}$ step ($k \geq 2$), we observe that, by definition of $\mathcal{F}_{k-1}$:
\[ \matr{R}^n\mathcal{F}_{k-1} = \matr{R}^n\mathcal{F}_{k-2} + \matr{R}^n<\vect{e}_k >_\R \ , \]
and by the above observations:
\[ \matr{R}^n\mathcal{F}_{k-2} = \matr{R}^{k-1}<\vect{e}_{k-1}, \vect{e}_{k-2}, \dots, \vect{e}_1 >_\R \ . \]
Hence, taking $\vect{e}_{n-k+1} \in \mathcal{F}_{k-1}$, it follows that:
\begin{align*}
-\epsilon\matr{R}^n\vect{e}_{n-k+1} & = \epsilon\matr{R}^{k-1}\big( p_{n-1}\vect{e}_1 + \cdots + p_{n-k+1}\vect{e}_{k-1} + p_{n-k}\vect{e}_{k} + \cdots + p_1\vect{e}_{n-1} + \epsilon\vect{e}_n \big) \\
 & \equiv \epsilon\matr{R}^{k-1} \big( p_{n-k}\vect{e}_{k} + \cdots + p_1\vect{e}_{n-1} + \epsilon\vect{e}_n \big) \mod \matr{R}^n\mathcal{F}_{k-2} \\
 & = \epsilon p_{n-k}\vect{e}_1 + \cdots + \epsilon p_1\vect{e}_{n-k} + \vect{e}_{n-k+1} \ .
\end{align*}
Comparing this expression with that obtained from application of $\matr{U}$:
\[ \matr{U}\vect{e}_{n-k+1} = u_{1,n-k+1}\vect{e}_1 + \cdots + u_{n-k,n-k+1}\vect{e}_{n-k} + \vect{e}_{n-k+1} \ , \]
we see that we have determined the $(n-k+1)^\matr{th}$ column of $\matr{S}$, for all $1 \leq k \leq n$.  Evidently, we have deduced that $u_{ij} = \epsilon p_{i-j}$, from which we conclude that $\matr{U} = \matr{S}$.

Now, we repeat the above, but for the matrix $-\epsilon\matr{R}^{-n} = \matr{L^{-1}U^{-1}}$ and the flag:
\[ \mathcal{\tilde{F}} \ : \ \vect{0} \subset \mathcal{\tilde{F}}_0 \subset \mathcal{\tilde{F}}_1 \subset \cdots \subset \mathcal{\tilde{F}}_{n-1} \cong \R^n \ , \quad \mathcal{\tilde{F}}_k := \, <\vect{e}_1, \dots, \vect{e}_{k+1}>_\R \ , \]
which is fixed by $\matr{U}^{-1}$.  It follows that $\vect{e}_{k+1} = \matr{R}^{-k}\vect{e}_1$ for all $0 \leq k \leq n-1$, from which it follows, by Cayley-Hamilton and (\ref{Palindromic}), that:
\[ \matr{R}^{-n}\vect{e}_1 = -p_{n-1}\vect{e}_n - \cdots - p_1\vect{e}_2 - \epsilon\vect{e}_1 \ . \]
We then deduce that $\matr{L}^{-1} = \matr{S^T}$, as was to be shown. \quad $\blacksquare$


For the next proposition, let $\omega := e^{\frac{2\pi i}{n}}$, and define the matrix $\Pi_{\epsilon} \in \matr{GL}_n\R$ by:
\begin{equation}\label{PiMatrix}
\Pi_{\epsilon} := \left[ \begin{array}{c | c} \vect{0} & \matr{I}_{n-1} \\ \hline \epsilon & \vect{0} \end{array} \right] \ .
\end{equation}
Then the characteristic polynomial of $\Pi_{\epsilon}$ is $x^n - \epsilon$, and hence, its eigenvalues are:
\[ \pi_k := \begin{cases}               \omega^k \ , & 0 \leq k \leq n-1 \ , \ n=2m \ , \\
                          \omega^{k+\frac{1}{2}} \ , & 0 \leq k \leq n-1 \ , \ n=2m+1 \ . \end{cases} \]

\begin{prop}\label{SignatureProp}
The eigenvalues of $\epsilon(\matr{S+S^T})$ are $p(\pi_0), \dots, p(\pi_{n-1})$.
\end{prop}

\prf Since $p(x)$ satisfies (\ref{Palindromic}), it is then evident that $\epsilon(\matr{S+S^T}) = p(\Pi_{\epsilon})$, and hence, $\matr{S+S^T}$ commutes with $\Pi_{\epsilon}$, so they may be diagonalized simultaneously. To each $\pi_k$, then, we let $\vect{v}_k$ denote the corresponding eigenvector of $\Pi_{\epsilon}$:
\begin{equation}\label{v_kVector}
\vect{v}_k := \frac{1}{n} \, (1, \pi_k, \pi_k^2, \dots, \pi_k^{n-1} )^\matr{T} \ , \quad 0 \leq k \leq n-1 \ ,
\end{equation}
It then follows that $\epsilon(\matr{S+S^T})\vect{v}_k = p(\pi_k)\vect{v}_k$ for each $k$. \quad $\blacksquare$

\begin{cor}\label{SignatureCor}
$\matr{S+S^T}$ has the same signature as the diagonal matrix \\ $\matr{diag} \, ( \epsilon p(\pi_0 ), \dots, \epsilon p(\pi_{n-1}) )$. In particular:
\begin{itemize}
\item $\matr{S+S^T}$ is positive definite \emph{iff} $\epsilon p(\pi_k) > 0$ for all $k$.
\item The number of zero eigenvalues is the number of common eigenvalues of $\matr{R}$ and $\Pi_{\epsilon}$. \quad $\blacksquare$
\end{itemize}
\end{cor}



As before, let us write $n=2m$ for even $n$, and $n=2m+1$ for odd $n$.  We recall that a complex number of unit norm is said to be \emph{unimodular}.
\begin{prop}\label{PosDefPolytope}(cf. \cite{CV3}, \cite{GIL2})
If $\matr{S+S^T}$ is positive definite, then the eigenvalues of $\matr{R}$ are unimodular. Moreover, the set of all $\matr{R}$ such that $\matr{S+S^T}$ is positive definite is in 1-1 correspondence with the bounded convex region of $\R^m$ defined by:
\[ \mathcal{P} := \bigcap_{k=0}^{m} \{ (p_1,\dots, p_m) \in \R^m \ | \ \epsilon p(\pi_k) > 0 \} \ . \]
\end{prop}

\prf $\matr{SS^{-T}}$ preserves the symmetric bilinear form $\matr{S+S^T}$ \cite{CV3}:
\[ \big( \matr{SS^{-T}} \big)^\matr{T} (\matr{S+S^T}) \big( \matr{SS^{-T}} \big) = \matr{S+S^T} \ . \]
Hence, $\matr{SS^{-T}}$ is orthogonal, and all eigenvalues of $\matr{SS^{-T}}$ are unimodular.  As a result, by Proposition \ref{NthRoot}, the eigenvalues of $\matr{R}$ must be unimodular, as well.

Next, we establish that the set of all $\matr{R}$ defined as in (\ref{RMatrix}), such that $\matr{S+S^T}$ is positive definite, is in 1-1 correspondence with $\mathcal{P}$:
\begin{itemize}
\item By Corollary \ref{SignatureCor}, $\epsilon p(\pi_k) > 0$ for all $k$, and thus, the entries $p_1, \dots, p_m$ of each such $\matr{R}$ determine a unique point $(p_1, \dots, p_m) \in \mathcal{P}$.
\item Conversely, given a point $P := (p_1, \dots, p_m) \in \mathcal{P}$, and defining $\matr{R}, \matr{S}$ according to (\ref{RMatrix}), (\ref{Palindromic}) and (\ref{StokesMatrix}) via the components of $P$, it follows by the definition of $\mathcal{P}$ and by Corollary \ref{SignatureCor} that $\matr{S+S^T}$ is positive definite, and hence, by the first assertion, all eigenvalues of $\matr{R}$ are unimodular.
\end{itemize}

Since the entries $p_1, \dots, p_m$ of $\matr{R}$ are the elementary symmetric polynomials of the eigenvalues of $\matr{R}$, all of which lie, by assumption, in the compact set $S^1$, $\mathcal{P}$ is consequently contained in the continuous image of a compact set, and hence, is bounded.

Lastly, as each inequality $\epsilon p(\pi_k) > 0$ defines a convex region of $\R^m$, and that the intersection of any collection of convex sets is convex, we see that $\mathcal{P}$ is convex. \ $\blacksquare$


Taking $\matr{S+S^T}$ non-degenerate, henceforth, we shall now consider the dependence of the signature $\sigma$ of $\matr{S+S^T}$ on the eigenvalues of $\matr{R}$, when $\matr{R}$ has only unimodular eigenvalues. By Proposition \ref{SignatureProp}\,\footnote{This was observed in \cite[p.27]{CV3} in the context of the general $tt^*$-equation.}, we note that $\sigma$ is constant with respect to any variation (within $S^1$) of an eigenvalue of $\matr{R}$ such that the eigenvalue does not pass through a root of $x^n-\epsilon$.  Hence, $\sigma$ is a function of only the number of eigenvalues of $\matr{R}$ between each root of $x^n - \epsilon$.  When $n=2m+1$, the conjugate symmetry of the eigenvalues implies that $\sigma$ is also a function of the number of eigenvalues in the arc $\{ e^{i\theta} \ | \ \theta \in [0, \tfrac{\pi}{2m+1}) \}$.  We now introduce some notation to assist in discussing this:
\begin{defn}
Assume $\matr{R}$ has only unimodular eigenvalues $e^{\pm i \theta_j}$, $1 \leq j \leq m$. (For $n=2m+1$, we do not include the guaranteed eigenvalue $z=1$ in this list.) When $n=2m$, the \emph{configuration $\rho$ of $\matr{R}$} is defined to be $\rho = (\rho_1, \dots, \rho_m) \in \Z_{\geq 0}^m$ such that $\rho_k := \# \left\{ j \ \vline \ \theta_j \in \big( \tfrac{(k-1)\pi}{m}, \tfrac{k\pi}{m} \big) \right\}$. For $n=2m+1$, the \emph{configuration $\rho$ of $\matr{R}$} is defined to be $\rho = (\rho_0, \dots, \rho_m) \in \Z_{\geq 0}^{m+1}$ such that:
\[ \rho_k := \left\{ \begin{array}{cl}
\# \left\{ j \ \vline \ \theta_j \in \big[ 0, \tfrac{\pi}{2m+1} \big) \right\} \ , & k=0 \ , \\
\# \left\{ j \ \vline \ \theta_j \in \big( \tfrac{(2k-1)\pi}{2m+1}, \tfrac{(2k+1)\pi}{2m+1} \big) \right\} \ , & 1 \leq k \leq m \ . 
\end{array}\right. \]
\upshape{Necessarily, the sum of the components of $\rho$ is always $m$, for all $n$, and we say that two matrices (for the same $n$) have the same configuration whenever their configuration sequences agree.}
\end{defn}





To find $\sigma$ in terms of $\rho$, we shall apply Descartes' Rule of Signs to a polynomial with only real roots uniquely derived from $p(x)$, and then relate this to a formula proved via an adaptation of the argument of \cite[Theorem 4.5]{BH}.  The result will then be, in principle, a formula to determine $\sigma$ from only the entries of $\matr{S}$.

\subsection{}

Recall that Descartes' Rule of Signs is the following classical result, whose proof we omit:

\begin{prop}(cf. \cite[Theorem 6.2d]{Henrici})\label{Descartes}
Let $p(x) \in \R[x]$ have degree $n \in \N$, with non-zero leading coefficient $a_n$, and let $\nu$ denote the number of sign changes in the sequence of non-zero coefficients of $p(x)$, starting with $a_n$ and listed in decreasing order of the power of $x$. If $r$ denotes the number of real, positive roots of $p(x)$, where each root is counted according to its algebraic multiplicity, then $\nu - r$ is even and non-negative.
\end{prop}

To refine this for polynomials with only real roots, we first prove:
\begin{lemma}
Let $p(x) = \sum_{k=0}^{n} a_k x^k \in \R[x]$ be non-zero, and let $\nu$ and $\mu$ be the number of sign changes in the decreasing sequence of non-zero coefficients of $p(x)$ and $p(-x)$, respectively. Then $\nu + \mu \leq n$, and equality holds \emph{iff} $a_k \neq 0$ for all $k$.
\end{lemma}

\prf Let $\sigma(a_k, a_{k-1})$ denote the number of sign changes in the 2-term sequence $(a_k, a_{k-1})$, allowing this to be 0 when at least one of $a_k$ or $a_{k-1}$ are 0. Then by definition of $\nu$, $\nu = \sum_{k=1}^{n} \sigma(a_k, a_{k-1})$. Now, by inspection:
\[ \chi_k := \sigma(a_k, a_{k-1}) + \sigma((-1)^k a_k , (-1)^{k-1} a_{k-1}) = \begin{cases} 1 \ , & a_k \neq 0 \ \textrm{and} \ a_{k-1} \neq 0 \ , \\
                                                                                  0 \ , & a_k = 0 \ \textrm{or} \ a_{k-1} = 0 \ , \end{cases} \]
so summing over all $k$, it follows that $\sum_{k=1}^{n} \chi_k = \mu + \nu$.  But the left-hand side is an $n$-term summation of 1s and 0s, so $\mu + \nu \leq n$, and $\mu + \nu = n$ \emph{iff} all $n$ terms of the sum are 1, \emph{iff} $a_k \neq 0$ for all $0 \leq k \leq n$. \quad $\blacksquare$

\begin{cor}\label{RealRoots}
If $p(x) \in \R[x]$ has only real roots, then $\nu=r$.  
\end{cor}

\prf Let $\nu$ and $\mu$ be defined as in the lemma, and let $r$ and $s$ be the number of positive roots of $p(x)$ and $p(-x)$, respectively. (Clearly, $s$ is the number of negative roots of $p(x)$.) Then by Descartes' Rule applied to both $p(x)$ and $p(-x)$, there are $\alpha, \beta \in \Z_{\geq 0}$ such that $\nu - r = 2\alpha$ and $\mu - s = 2\beta$. 

First, assume that $p(x)$ has only non-zero real roots, so that $n = r + s$. Then $n = \nu + \mu - 2(\alpha + \beta)$, so by the lemma and non-negativity of $\alpha$ and $\beta$, $\alpha = \beta = 0$.

Now suppose that $p(x)$ has only real roots, and $t$ of them zero.  Then $p(x)x^{-t}$ has only non-zero real roots, and the number of its positive roots is also $r$, so by the previous assertion, if $\nu'$ is the number of sign changes of $p(x)x^{-t}$, then $\nu' = r$.  But evidently, $\nu=\nu'$, which proves the assertion. \quad $\blacksquare$

Now, consider the following: for any $p(x) \in \R[x]$ satisfying (\ref{Palindromic}), it may be shown by induction that there is a unique monic $\tilde{p}(x) \in \R[x]$ such that:
\begin{equation}\label{ReducedPoly}
\begin{cases}
     p(x) = x^m \tilde{p} \big( x + \tfrac{1}{x} \big) \ , & n=2m   \ , \\
p(x) = (x-1) x^m \tilde{p}\big( x + \tfrac{1}{x} \big) \ , & n=2m+1 \ , \end{cases}
\end{equation}
(For $n=2m+1$, we note that $p(1)=0$ by (\ref{Palindromic}), so after factoring $p(x) = (x-1)q(x)$, we see that $q(x)$ satisfies (\ref{Palindromic}), and thus, the even case factorization applies.)

As all eigenvalues of $\matr{R}$ are unimodular, $\tilde{p}(x)$ has one root $2\cos\theta_j$ for each conjugate pair of roots $e^{\pm i\theta_j}$ of $p(x)$. This motivates the following:
\begin{prop}\label{SignChangeVector}
Given $\tilde{p}(x)$ as in (\ref{ReducedPoly}), let $\tilde{p}^{[0]}(x) := \tilde{p}(x+2)$ for all $n$, and:
\[ \tilde{p}^{[k]}(x) := \begin{cases}
        \tilde{p}\big( x + 2\cos\tfrac{k\pi}{m} \big) \ , & 1 \leq k \leq m   \ , \ n=2m   \ , \\
\tilde{p}\big( x + 2\cos\tfrac{(2k-1)\pi}{2m+1} \big) \ , & 1 \leq k \leq m+1 \ , \ n=2m+1 \ . \end{cases} \]
Denote by $\nu_k$ the number of sign changes in the sequence of non-zero coefficients of $\tilde{p}^{[k]}(x)$, as in Descartes' Rule.  Then the configuration $\rho$ of $\matr{R}$ satisfies:
\[ \nu_k - \nu_{k-1} = \begin{cases}
    \rho_k \ , & 1 \leq k \leq m   \ , \ n=2m   \ , \\
\rho_{k-1} \ , & 1 \leq k \leq m+1 \ , \ n=2m+1 \ . \end{cases} \]
Conversely, $\nu_k$ is given by:
\[ \nu_k = \begin{cases}
  \sum_{j=1}^{k} \rho_j \ , & 0 \leq k \leq m   \ , \ n=2m   \ , \\
\vphantom{\Big( \Big)} \sum_{j=0}^{k-1} \rho_j \ , & 0 \leq k \leq m+1 \ , \ n=2m+1 \ . \end{cases} \]
\end{prop}

\prf Evidently, when $n=2m$, $\rho_k$ is the number of roots of $\tilde{p}(x)$ in the interval $\big( 2\cos\tfrac{k\pi}{m}, 2\cos\tfrac{(k-1)\pi}{m} \big)$, and when $n=2m+1$, $\rho_k$ is the number of roots of $\tilde{p}(x)$ in:
\[ \begin{cases}
                 (2\cos\tfrac{\pi}{2m+1}, 2]                           \ , & k=0 \ , \\
\Big( 2\cos\tfrac{(2k+1)\pi}{2m+1}, 2\cos\tfrac{(2k-1)\pi}{2m+1} \Big) \ , & 1 \leq k \leq m \ . \end{cases} \]
On the other hand, by construction of the $\tilde{p}^{[k]}(x)$, the number of positive roots $r_k$ of $\tilde{p}^{[k]}(x)$ is the same as the number of roots of $\tilde{p}(x)$ strictly greater than $2\cos\tfrac{k\pi}{m}$ when $n=2m$, and when $n=2m+1$, it is the same as the number of roots of $\tilde{p}(x)$ strictly greater than $2$, for $k=0$, or $2\cos\tfrac{(2k-1)\pi}{2m+1}$ for all other $k$.  Moreover, all $m$ roots of $\tilde{p}(x)$ are real, and thus, $\nu_k = r_k$ by Corollary \ref{RealRoots}.  Applying Descartes' Rule to $\tilde{p}^{[k]}(x)$ and $\tilde{p}^{[k-1]}(x)$, it then follows that:
\[ \nu_k - \nu_{k-1} = \begin{cases}
    \rho_k \ , & 1 \leq k \leq m   \ , \ n=2m   \ , \\
\rho_{k-1} \ , & 1 \leq k \leq m+1 \ , \ n=2m+1 \ . \end{cases} \]
Conversely, given $\rho$, it follows from the above that, for all $k$:
\[ \nu_k = \begin{cases}
  \sum_{j=1}^{k} \rho_j \ , & 0 \leq k \leq m   \ , \ n=2m   \ , \\
\sum_{j=0}^{k-1} \rho_j \ , & 0 \leq k \leq m+1 \ , \ n=2m+1 \ . \quad \blacksquare \end{cases} \]

For notational convenience, we shall always denote the sequence of the number of sign changes of $\tilde{p}^{[k]}(x)$ by:
\[ \nu := \begin{cases}
(0, \nu_1, \dots, \nu_{m-1}, m) \ , & n=2m   \ , \\
    (0, \nu_1, \dots, \nu_m, m) \ , & n=2m+1 \ . \end{cases} \]

\begin{rmrk}\upshape
For the matrix $\matr{R}$ with characteristic polynomial $p(x) = x^{2m}+1$, $\matr{S+S^T} = 2\matr{I}_{2m}$, and the configuration is $\rho = (1,1,\dots,1)$, which corresponds to the sequence $\nu = (0,1,2,\dots,m-1,m)$.  Hence, by connectedness of $\mathcal{P}$ in Proposition \ref{PosDefPolytope}, for any $\matr{R}$ with only unimodular eigenvalues when $n=2m$, $\matr{S+S^T} > 0$ \emph{iff} its configuration is $(1,1,\dots,1)$, which is \emph{iff} its sequence of sign-change numbers is $(0,1,2\dots,m-1,m)$.  Similarly, for the matrix $\matr{R}$ with characteristic polynomial $p(x) = x^{2m+1} - 1$, we have $\matr{S+S^T} = 2\matr{I}_{2m+1}$, $\rho=(0,1,1,\dots,1)$, and $\nu = (0,0,1,2,\dots,m-1,m)$, and this is the only configuration for which $\matr{S+S^T} > 0$.

This observation has the interpretation (cf. Corollary 4.7 of \cite{BH}) that $\matr{S+S^T} > 0$ \emph{iff} the eigenvalues $e^{\pm i\theta_j}$ of $\matr{R}$ \emph{interlace} with the roots $\pi_k$ of $\epsilon$ (including the guaranteed root $z = e^{i \, 0}$ when $n=2m+1$):
\[ \left\{ \begin{array}{cl}
0 < \theta_1 < \tfrac{\pi}{m} < \theta_2 < \tfrac{2\pi}{m} < \cdots < \tfrac{(m-1)\pi}{m} < \theta_m < \pi \ , & n=2m \ , \\
\tfrac{-\pi}{2m+1} < 0 < \tfrac{\pi}{2m+1} < \theta_1 < \tfrac{3\pi}{2m+1} < \cdots < \tfrac{(2m-1)\pi}{2m+1} < \theta_m < \pi \ , & n=2m+1 \ .
\end{array}\right. \]
\end{rmrk}

\subsection{}

Inspired by Sections 3 and 4 of \cite{BH}, we prove a formula for $\sigma$ using the sequence $\nu$ of sign-change numbers of the $\tilde{p}^{[k]}(x)$.  Before proving the formula, we first note:
\begin{enumerate}
\item Since the characteristic polynomial $p(x)$ of $\matr{R}$ satisfies (\ref{Palindromic}), it follows that $\det(x\matr{I}_n - \matr{R^{-T}})=p(x)$, as well.
\item Letting $\matr{D} := \matr{I}_n - \Pi_{\epsilon}\matr{R^T}$, we remark that $\matr{D}$ has rank 1, and for all $\vect{x} \in \C^n$, $\matr{D}\vect{x} = \big( \vect{e}_n^*(\matr{S+S^T})\vect{x} \big) \vect{e}_n$.  Note, as well, that $(\Pi_{\epsilon}\matr{R^T})^2=\matr{I}_n$.
\item \cite{BH} If a rank one $n \times n$ matrix $\matr{M}$ acts on $\C^n$ as $\matr{M}\vect{x} = w(\vect{x})\vect{u}$ for some linear form $w$ and for some $\vect{u} \in \C^n$, then $\det(\matr{I}_n + \matr{M}) = 1 + w(\vect{u})$.
\item Letting $\vect{v}_k$ be defined as in (\ref{v_kVector}), we note that $\sum_{k=0}^{n-1} \pi_k \vect{v}_k = \Pi_{\epsilon}\vect{e}_1 = \epsilon \, \vect{e}_n$.
\end{enumerate}

We now prove what is essentially a special case of Theorem 4.5 of \cite{BH}:
\begin{prop}\label{SignFormula}
For $\matr{R}$ with only unimodular eigenvalues such that $\matr{S+S^T}$ is non-degenerate, denote the eigenvalues of $\matr{R}$ as $z_k := e^{2\pi i \theta_k}$, where $\theta_k \in [0,1)$, for $0 \leq k \leq n-1$.  When $n=2m+1$, we take $z_m = 1$, so $\theta_m = 0$.  If $n_j := \# \{ k \ | \ \theta_k < \matr{arg}\pi_j \}$ for each $0 \leq j \leq n-1$, then $sgn(p(\pi_j)) = (-1)^{n_j - j}$.
\end{prop}

\prf We adapt the method of proof of Theorem 4.5 of \cite{BH} as follows: Expanding $p(x)$ and $x^n - \epsilon$ into their complex linear factors, we use the above remarks to obtain:
\begin{align*}
\prod_{k=0}^{n-1} (z_k - x)(\pi_k - x)^{-1} & = \det(\matr{R} - x\matr{I}_n)\det(\Pi_{\epsilon} - x\matr{I}_n)^{-1} \\
 & = \det(\matr{R^{-T}} - x\matr{I}_n)\det(\Pi_{\epsilon} - x\matr{I}_n)^{-1} \\
 & = \det\left( (\matr{R^{-T}}\Pi_{\epsilon}^{-1} - x\Pi_{\epsilon}^{-1})(\matr{I}_n - x\Pi_{\epsilon}^{-1})^{-1} \right) \\
 & = \det\left( (-\matr{D} + \matr{I}_n - x\Pi_{\epsilon}^{-1})(\matr{I}_n - x\Pi_{\epsilon}^{-1})^{-1} \right) \\
 & = \det\left( \matr{I}_n - \matr{D}(\matr{I}_n - x\Pi_{\epsilon}^{-1})^{-1} \right) \\
 & = 1 - \vect{e}_n^*(\matr{S+S^T})(\matr{I}_n - x\Pi_{\epsilon}^{-1})^{-1}\vect{e}_n \\
 & = 1 - \vect{e}_n^*(\matr{S+S^T})(\Pi_{\epsilon} - x\matr{I}_n)^{-1}\Pi_{\epsilon}\vect{e}_n \ .
\end{align*}
Using the expression for $\vect{e}_n$ in terms of the $\vect{v}_k$ then yields:
\[ \prod_{k=0}^{n-1} (z_k - x)(\pi_k - x)^{-1} = 1 - \sum_{k,j=0}^{n-1} \frac{\pi_j^2 \overline{\pi}_k}{\pi_j - x} \vect{v}_k^* (\matr{S+S^T})\vect{v}_j = 1 - \sum_{j=0}^{n-1} \frac{\pi_j}{\pi_j - x} \vect{v}_j^* (\matr{S+S^T})\vect{v}_j \ . \]
Taking the residue at $x=\pi_j$, and inserting\footnote{Note that $z_k^{\frac{1}{2}}z_{n-k}^{\frac{1}{2}} = -1$ for all $k$, since $\theta_k \in [0,1)$ and the $z_k$ come in conjugate pairs, except for $z_m=1$ when $n=2m+1$. A similar statement holds for the $\pi_k$.} $\epsilon = i \prod_{k=0}^{n-1} (\pi_k)^{\frac{1}{2}} \, z_k^{-\frac{1}{2}}$, we find that:
\begin{align*}
\epsilon \, \vect{v}_j^* (\matr{S+S^T})\vect{v}_j & = - \epsilon (z_j - \pi_j)\pi_j^{-1} \prod_{k \neq j} (z_k - \pi_j)(\pi_k - \pi_j)^{-1} \\
 & = - \left( i\prod_{k=0}^{n-1} \pi_k^{\frac{1}{2}} \, z_k^{-\frac{1}{2}} \right) (z_k - \pi_j)\pi_j^{-1} \prod_{k \neq j} (z_j - \pi_j)(\pi_k - \pi_j)^{-1} \\
 & = - i \left( \frac{z_j^\frac{1}{2}}{\pi_j^\frac{1}{2}} - \frac{\pi_j^\frac{1}{2}}{z_j^\frac{1}{2}} \right) \prod_{k \neq j} \left( \frac{z_k^\frac{1}{2}}{\pi_j^\frac{1}{2}} - \frac{\pi_j^\frac{1}{2}}{z_k^\frac{1}{2}} \right) \left( \frac{\pi_k^\frac{1}{2}}{\pi_j^\frac{1}{2}} - \frac{\pi_j^\frac{1}{2}}{\pi_k^\frac{1}{2}} \right)^{-1} \\
 & = 2 \sin\pi(\theta_j - \matr{arg}\pi_j) \prod_{k \neq j} \frac{ \sin\pi(\theta_k - \matr{arg}\pi_j) }{ \sin(\matr{arg}\pi_k - \matr{arg}\pi_j) } \ .
\end{align*}
The sign of the denominator is $(-1)^j$, by inspection, and the sign of the numerator is $(-1)^{n_j}$, by definition of $n_j$. Thus, $sgn(p(\pi_j)) \! = \! (-1)^{n_j - j}$, by Proposition \ref{SignatureProp}. \ $\blacksquare$

\begin{cor}\label{TheFormula}
Let $\nu$ be the sequence of sign-change numbers of $\tilde{p}^{[k]}(x)$, and let $\matr{S+S^T}$ have $n_+$ positive and and $n_-$ negative eigenvalues.  Then for $n=2m$:
\[ \begin{cases}
     n_j = \nu_j      \ , & 0 \leq j \leq m   \ , \\
n_{2m-j} = 2m - \nu_j \ , & 1 \leq j \leq m-1 \ , \end{cases} \]
and for $n=2m+1$:
\[ \begin{cases}
     n_j - 1 = \nu_{j+1}      \ , & 0 \leq j \leq m \ , \\
n_{2m-j} - 1 = 2m - \nu_{j+1} \ , & 0 \leq j \leq m-1 \ . \end{cases} \]
Consequently, for all $j$:
\[ sgn(p(\pi_j)) = \begin{cases}
        (-1)^{\nu_j - j} \ , & n=2m   \ , \\
(-1)^{\nu_{j+1} - (j+1)} \ , & n=2m+1 \ , \end{cases} \]
and thus, for all $n$:
\[ n_+ - m - 1 = \begin{cases}
\sum_{j=1}^{m-1} (-1)^{\nu_j - j} \ , & n=2m   \ , \\
 -\sum_{j=1}^{m} (-1)^{\nu_j - j} \ , & n=2m+1 \ . \end{cases} \]
\end{cor}

\prf The relations between $n_j$ and $\nu_j$ follow from the definition of $n_j$ and from Proposition \ref{SignChangeVector}.  Note that, when $n=2m+1$, $n_j \geq 1$ for all $j$ due to the guaranteed root $z_m=1$ of $p(x)$.  By Proposition \ref{SignFormula}, the formula for $sgn(p(\pi_j))$ then follows immediately.  Since $n_+ + n_- = n$ whenever $\matr{S+S^T}$ is non-degenerate, it follows by the above formula for $sgn(p(\pi_j))$ that:
\[ \epsilon (2n_+ - n) = \epsilon (n_+ - n_-) = \begin{cases}
(-1)^{\nu_0} + (-1)^{\nu_m-m} + 2 \ \sum_{j=1}^{m-1} (-1)^{\nu_j - j} \ , & n=2m   \ , \\
         (-1)^{\nu_{m+1}-(m+1)} + 2 \ \sum_{j=1}^{m} (-1)^{\nu_j - j} \ , & n=2m+1 \ . \end{cases} \]
When $n=2m$, $\nu_j=j$ for $j=0$ and $j=m$, and hence:
\[ n_+ = m + 1 + \sum_{j=1}^{m-1} (-1)^{\nu_j - j} \ . \]
When $n=2m+1$, $\nu_{m+1}=m$ for $j=m$, and hence:
\[ n_+ = m + 1 - \sum_{j=1}^{m} (-1)^{\nu_j - j} \ . \quad \blacksquare \]

\begin{rmrk}\upshape

We now provide some sample calculations. When $n=4$ (cf. \cite{GIL2}), the relation between $\sigma = ( n_+ , n_- )$ and $\nu=(0,\nu_1,2)$ is:
\begin{center}
\begin{tabular}{ c | c | c | c }
 $\nu_1$ &    0    &    1    &    2    \\ \hline
$\sigma$ & $(2,2)$ & $(4,0)$ & $(2,2)$
\end{tabular}
\end{center}
When $n=6$, the signatures may be tabulated as follows:
\begin{center}
\begin{tabular}{ c | c | c | c | c }
$\nu_1 \ \diagdown \ \nu_2$ &    0     &    1     &    2     &   3     \\ \hline
             0              &  $(4,2)$ &  $(2,4)$ &  $(4,2)$ & $(2,4)$ \\ \hline
             1              & $\times$ &  $(4,2)$ &  $(6,0)$ & $(4,2)$ \\ \hline
             2              & $\times$ & $\times$ &  $(4,2)$ & $(2,4)$ \\ \hline
             3              & $\times$ & $\times$ & $\times$ & $(4,2)$  
\end{tabular}
\end{center}
\end{rmrk}

\section{Application to the $tt^*$-Toda equation}

Now we shall apply the results of the previous section to the symmetrized Stokes matrices of the $tt^*$-Toda equation, which were calculated in \cite{GIL1, GIL2, MoXX}.  Let us consider the pole at infinity (of order 2) of Equation (\ref{RadialCompatEqtn}), for arbitrary $n+1$.  In Section 4 of \cite{GIL1}, the case $n+1=4$ was treated in detail.  The same method applies for any $n \geq 3$, so we shall just summarize the results briefly.

Recall that we wish to determine the Stokes data at $\zeta = \infty$ for the ODE:
\begin{equation}\label{ZetaEqtn}
\Psi_\zeta = \ \left( -\tfrac{1}{\zeta^2} \matr{W} - \tfrac{1}{\zeta} x\matr{w}_x + x^2\matr{W^T} \right) \Psi \ ,
\end{equation}
where $w$ and $\matr{W}$ are defined before (\ref{RadialCompatEqtn}). If $\eta := \zeta^{-1}$, then we may re-write this as:
\[ \Psi_\eta = \bigg( -\tfrac{1}{\eta^2} x^2 \matr{W^T} + O\big(\tfrac{1}{\eta}\big) \bigg) \Psi \ . \]
Letting $\omega := e^{\frac{2\pi i}{n+1}}$, $\matr{d}_{n+1} := \matr{diag} \, (1, \omega, \dots, \omega^n )$, and $\Omega := \big( \omega^{ij} \big)_{i,j=0}^{n}$, we may use the matrix $\matr{P}_{\infty} := \matr{diag} \, ( e^{w_0}, \dots, e^{w_n} ) \, \Omega^{-1}$ to diagonalize $\matr{W^T}$ as:
\[ \matr{W^T} = \matr{P}_{\infty} \matr{d}_{n+1} \matr{P}_{\infty}^{-1} \ . \]
Then by Proposition 1.1 of \cite{FIKN}, and reverting back to $\zeta$, we see that there exists a unique formal solution $\Psi_f^{\infty}$ of (\ref{ZetaEqtn}) of the form:
\[ \Psi_f^{\infty} = \matr{P}_{\infty} \left( \matr{I}_{n+1} + \sum_{j \geq 1} \psi_j^{\infty} \zeta^{-j} \right) e^{\Lambda_0\log\eta + x^2\zeta\matr{d}_{n+1}} \ . \]
It may then be verified, by direct substitution into (\ref{ZetaEqtn}), that $\Lambda_0 = 0$.  By Theorem 1.4 of \cite{FIKN}, there is then a unique holomorphic solution $\Psi$ to (\ref{ZetaEqtn}), with asymptotic expansion $\Psi_f^{\infty}$, on any Stokes sector based at $\zeta=\infty$.

There are $2(n+1)$ Stokes rays, given by all $\zeta \in \C^*$ satisfying:
\begin{equation}\label{StokesRays}
\cos\big(\matr{arg}\zeta + \matr{arg}(\omega^j - \omega^k) \big) = 0 \ , 
\end{equation}
where $\omega := e^{ \frac{2\pi i}{n+1} }$.  As fundamental Stokes sectors, when $n+1=2m$, we take:
\[ \Omega_1^{\infty} = \{ \zeta \in \C^* \ | \ -\tfrac{\pi}{2} < \matr{arg}\zeta < \tfrac{\pi}{2} + \tfrac{\pi}{n+1} \} \ , \]
\[ \Omega_2^{\infty} = \{ \zeta \in \C^* \ | \  \tfrac{\pi}{2} < \matr{arg}\zeta < \tfrac{3\pi}{2} + \tfrac{\pi}{n+1} \} \ , \]
and when $n+1=2m+1$, we take:
\begin{gather*}
\Omega_1^{\infty} = \{ \zeta \in \C^* \ | \ -\tfrac{\pi}{2} - \tfrac{\pi}{2(n+1)} < \matr{arg}\zeta < \tfrac{\pi}{2} + \tfrac{\pi}{2(n+1)} \} \ , \\
\Omega_2^{\infty} = \{ \zeta \in \C^* \ | \  \tfrac{\pi}{2} - \tfrac{\pi}{2(n+1)} < \matr{arg}\zeta < \tfrac{3\pi}{2} + \tfrac{\pi}{2(n+1)} \} \ .
\end{gather*}

The Stokes matrix $\matr{S}_1^{\infty}$ is defined by $\Psi_2^{\infty} = \Psi_1^{\infty}\matr{S}_1^{\infty}$, where $\Psi_1^{\infty}$ is the canonical solution with prescribed asymptotics on $\Omega_1^{\infty}$, and where the analytic continuation of $\Psi_1^{\infty}$ to $\Omega_2^{\infty}$ is taken in the positive direction.

Letting $\Pi := \left( \begin{smallmatrix} \vect{0} & \matr{I}_n \\ 1 & \vect{0} \end{smallmatrix} \right)$, and using the symmetries of (\ref{RadialCompatEqtn}), as in Section 4 of \cite{GIL1}, we find that:
\[ \matr{S}_1^{\infty} = \left\{ \begin{array}{cl}
\big( \matr{Q}_1^{\infty} \matr{Q}_{1\frac{1}{n+1}}^{\infty} \Pi \big)^m \, \Pi^{-m} \ , & n+1=2m \ , \\
\big( \matr{Q}_1^{\infty} \matr{Q}_{1\frac{1}{n+1}}^{\infty} \Pi \big)^m \, \matr{Q}_{1}^{\infty} \Pi^{-m} \ , & n+1=2m+1 \ , \end{array}\right. \]
and for all $n$, the inverse of the monodromy of $\Psi_1^{\infty}$ is:
\[ \matr{S}_1^{\infty} \matr{S}_2^{\infty} = \big( \matr{Q}_1^{\infty} \matr{Q}_{1\frac{1}{n+1}}^{\infty} \Pi \big)^{n+1} \ . \]
Here, the matrices $\matr{Q}_k^{\infty}$ are the $\llq$Stokes factors" of $\matr{S}_1^{\infty}$ and $\matr{S}_2^{\infty}$, defined with respect to the Stokes sectors $\Omega_{k+1}^{\infty} = e^{k\pi i}\Omega_1^{\infty}$ for all $k \in \tfrac{1}{n+1}\Z$ (i.e. $\Psi_{k+\frac{1}{n+1}}^{\infty} = \Psi_k^{\infty}\matr{Q}_k^{\infty}$).

As in Section 5 of \cite{GIL2}, we may convert to real matrices $\matr{\tilde{S}}_k^{\infty}$ and $\matr{\tilde{Q}}_k^{\infty}$ by conjugating by\footnote{For $n+1=2m$, $r=\tfrac{1}{2}$, and for $n+1=2m+1$, $r=m+1$.} $\big( \matr{diag} \, ( 1, \omega, \dots, \omega^n ) \big)^r$. Taking $\varepsilon := (-1)^{n+1}$ and:
\begin{equation}\label{RealRMatrix}
\matr{R}_{\varepsilon} := \begin{bmatrix} \vect{0} & \matr{I}_n \\ -\varepsilon & \vect{0} \end{bmatrix} \ , \quad
\mathcal{R} := \matr{\tilde{Q}}_1^{\infty} \matr{\tilde{Q}}_{1\frac{1}{n+1}}^{\infty} \matr{R}_{\varepsilon}  \ ,
\end{equation}
\begin{equation}\label{Pseudosymplectic}
\mathcal{J} := \left\{ \begin{array}{cl}
\matr{R}_{\varepsilon}^m = \left( \begin{smallmatrix} \vect{0} & \matr{I}_m \\ -\matr{I}_m & \vect{0} \end{smallmatrix} \right) \ , & n+1=2m \ , \\
\matr{R}_{\varepsilon}^m \big( \matr{\tilde{Q}}_1^{\infty} \big)^{-1} = \Pi^m \big( \matr{\tilde{Q}}_1^{\infty} \big)^{-1} \ , & n+1=2m+1 \ ,
\end{array}\right.
\end{equation}
we then obtain:
\begin{equation}\label{RealStokesMatrix}
\mathcal{S} := \matr{\tilde{S}}_1^{\infty} = \mathcal{R}^m\mathcal{J}^{-1} \ , \quad \mathcal{S}\mathcal{S}^{-\matr{T}} = \matr{\tilde{S}}_1^{\infty} \matr{\tilde{S}}_2^{\infty} = -\varepsilon \mathcal{R}^{n+1} \ .
\end{equation}

We are now almost in the situation of Section 2 of this article. Due to the choice of formal solutions made in \cite{GIL1}, the Stokes matrices $\matr{\tilde{S}}_k^{\infty}$ are elements of the group:
\[ \{ \matr{A} \in \matr{SL}_{n+1}\R \ | \ \matr{A^T}\mathcal{J}\matr{A} = \mathcal{J} \} \ , \]
but the $\llq$idealized" Stokes matrices of Section 2 are not, in general.  Therefore, to be able to apply the obtained results to $\mathcal{S} + \mathcal{S}^\matr{T}$, a further transformation is required.

To find the correct transformation, we need explicit expressions for the matrices $\tilde{\matr{Q}}_1^{\infty}$ and $\tilde{\matr{Q}}_{1\frac{1}{n+1}}^{\infty}$. Their non-zero entries can be deduced from:
\begin{lemma}
The diagonal entries of $\tilde{\matr{Q}}_1^{\infty}$ and $\tilde{\matr{Q}}_{1\frac{1}{n+1}}^{\infty}$ are $1$, and the other entries satisfy the rule (for $1 \leq i \neq j \leq n+1$):
\[ (n+1 = 2m) \left\{ \begin{array}{rcl} 
\matr{arg}(\omega^{i-1}-\omega^{j-1}) \neq \frac{n\pi}{n+1} \mod 2\pi & \Rightarrow & (\tilde{\matr{Q}}_1^{\infty})_{i,j} = 0 \ , \\
\matr{arg}(\omega^{i-1}-\omega^{j-1}) \neq \frac{(n-1)\pi}{n+1} \mod 2\pi & \Rightarrow & (\tilde{\matr{Q}}_{1\frac{1}{n+1}}^{\infty})_{i,j} = 0 \ . \end{array} \right. \]
\[ (n+1 = 2m+1) \left\{ \begin{array}{rcl} 
\matr{arg}(\omega^{i-1}-\omega^{j-1}) \neq \frac{(2n+1)\pi}{2(n+1)} \mod 2\pi & \Rightarrow & (\tilde{\matr{Q}}_1^{\infty})_{i,j} = 0 \ , \\
\matr{arg}(\omega^{i-1}-\omega^{j-1}) \neq \frac{(2n-1)\pi}{2(n+1)} \mod 2\pi & \Rightarrow & (\tilde{\matr{Q}}_{1\frac{1}{n+1}}^{\infty})_{i,j} = 0 \ . \end{array} \right. \]
\end{lemma}

\prf We follow the proof of Lemma 4.4 of \cite{GIL1}. For the complex Stokes factors $\matr{Q}_k^{\infty}$, $k \in \tfrac{1}{n+1}\Z$, we have:
\[ \matr{Q}_k^{\infty} = \lim_{\zeta \rightarrow \infty} (\Psi_k^{\infty})^{-1} \Psi_{k+\frac{1}{n+1}}^{\infty} = \lim_{\zeta \rightarrow \infty} e^{-\zeta x^2 \matr{d}_{n+1}} \big( \matr{I}_{n+1} + O(\tfrac{1}{\zeta}) \big) e^{\zeta x^2 \matr{d}_{n+1}} \ , \]
and hence, $(\matr{Q}_k^{\infty})_{ii} = 1$ for all $i$.  On the other hand, for $(i,j)$ such that $1 \leq i \neq j \leq n+1$, the entry $(\matr{Q}_k^{\infty})_{ij} = 0$ so long as there is a path $\zeta_t \rightarrow \infty$ in $\Omega_k \cap \Omega_{k+\frac{1}{n+1}}$ such that $\matr{Re} \, \zeta_t(\omega^{j-1} - \omega^{i-1}) \leq 0$. Since $\Omega_k \cap \Omega_{k+\frac{1}{n+1}}$ is a sector of angle $\pi$, it follows that $(\matr{Q}_k^{\infty})_{ij}$ is necessarily zero only if $(\omega^{j-1} - \omega^{i-1}) \, \Omega_k \cap \Omega_{k+\frac{1}{n+1}}$ overlaps with the closed half-plane $\{ \matr{Re} \, \zeta \leq 0 \}$. But this occurs \emph{iff}:
\[ \matr{arg}(\omega^{i-1} - \omega^{j-1}) \neq \begin{cases}
\tfrac{(2n+1-(n+1)k)\pi}{n+1} \mod 2\pi \ , & n+1 = 2m \ , \\
\tfrac{(4n+3-2(n+1)k)\pi}{2(n+1)} \mod 2\pi \ , & n+1 = 2m+1 \ . \end{cases} \]
It then follows by the definition of the $\tilde{\matr{Q}}_k^{(\infty)}$ that their entries satisfy the same conditions, and by substituting $k=1$ and $k=1\tfrac{1}{n+1}$, the assertion follows. \quad $\blacksquare$ \\

Consequently, when $n+1=2m$, the entries must satisfy are:
\[ \left\{ \begin{array}{rcl} 
\matr{arg}(\omega^{i-1}-\omega^{j-1})/\frac{\pi}{2m} \neq 2m-1 \mod 4m & \Rightarrow & (\tilde{Q}_1^{\infty})_{i,j} = 0 \ , \\
\matr{arg}(\omega^{i-1}-\omega^{j-1})/\frac{\pi}{2m} \neq 2m-2 \mod 4m & \Rightarrow & (\tilde{Q}_{1\frac{1}{n+1}}^{\infty})_{i,j} = 0 \ , \end{array} \right. \]
and when $n+1=2m+1$, we instead have:
\[ \left\{ \begin{array}{rcl} 
\matr{arg}(\omega^{i-1}-\omega^{j-1})/\frac{\pi}{2(2m+1)} \neq 4m+1 \mod 8m+4 & \Rightarrow & (\tilde{Q}_1^{\infty})_{i,j} = 0 \ , \\
\matr{arg}(\omega^{i-1}-\omega^{j-1})/\frac{\pi}{2(2m+1)} \neq 4m-1 \mod 8m+4 & \Rightarrow & (\tilde{Q}_{1\frac{1}{n+1}}^{\infty})_{i,j} = 0 \ . \end{array} \right. \]
Hence, the potentially non-zero entries are:
\begin{itemize}
\item \underline{$n+1=2m$}: $(\tilde{Q}_1^{\infty})_{m-k,1+k}$, $(\tilde{Q}_1^{\infty})_{m+1+k,2m-k}$,  $(\tilde{Q}_{1\frac{1}{n+1}}^{\infty})_{m-1-k,1+k}$ and \\ $(\tilde{Q}_{1\frac{1}{n+1}}^{\infty})_{m+k,2m-k}$, for $k \geq 0$ stopping before the diagonal.
\item \underline{$n+1=2m+1$}: $(\tilde{Q}_1^{\infty})_{m+1-k,1+k}$, $(\tilde{Q}_1^{\infty})_{m+2+k,2m+1-k}$, $(\tilde{Q}_{1\frac{1}{n+1}}^{\infty})_{m-k,1+k}$, and $(\tilde{Q}_{1\frac{1}{n+1}}^{\infty})_{m+1+k,2m+1-k}$ for $k \geq 0$ stopping before the diagonal.
\end{itemize}
Taking into account all of the symmetry conditions (see Section 5 of \cite{GIL2}), it can be deduced when $n+1=2m$ that for all $k \geq 0$:
\[ \begin{cases} (\tilde{Q}_1^{\infty})_{m-k,1+k} + (\tilde{Q}_1^{\infty})_{m+1+k,2m-k} = 0 \ , & \\
(\tilde{Q}_{1\frac{1}{n+1}}^{\infty})_{m-1-k,1+k} + (\tilde{Q}_{1\frac{1}{n+1}}^{\infty})_{m+1+k,2m-1-k} = 0 \ . & \end{cases} \]
For convenience of notation, let us define $p_m := (\tilde{Q}_{1\frac{1}{n+1}}^{\infty})_{m,2m}$ and:
\begin{equation}\label{EvenCaseEntries}
-p_{m-2k-2} := (\tilde{Q}_{1\frac{1}{n+1}}^{\infty})_{m-1-k,1+k} \ , \quad -p_{m-2k-1} := (\tilde{Q}_1^{\infty})_{m-k,1+k} \ .
\end{equation}
Then $\tilde{\matr{Q}}_1^{\infty}$ and $\tilde{\matr{Q}}_{1\frac{1}{n+1}}^{\infty}$ are the block-matrices:
\[ \tilde{\matr{Q}}_1^{\infty} = 
\left[ \begin{array}{c | c} 
\matr{\tilde{L}_1}^{\infty} & \vect{0} \vphantom{\bigg( \bigg)} \\ \hline 
\vect{0} & \vphantom{\bigg( \bigg)} \big(\matr{\tilde{L}_1}^{\infty}\big)^{-\matr{T}}
\end{array}\right] \ , \quad 
\tilde{\matr{Q}}_{1\frac{1}{n+1}}^{\infty} = 
\left[ \begin{array}{c | c} 
\matr{\tilde{L}}_{1\frac{1}{n+1}}^{\infty} & p_m \matr{E}_{m,m} \vphantom{\bigg( \big))} \\ \hline
\vect{0} \vphantom{\bigg( \bigg)} & \big(\matr{\tilde{L}}_{1\frac{1}{n+1}}^{\infty}\big)^{-\matr{T}} 
\end{array}\right] \ , \]
where:
\[ \matr{\tilde{L}}_1^{\infty} = \matr{I}_m - \sum_{k=0}^{\ell_1-1} p_{m-2k-1} \matr{E}_{m-k,1+k} \ , \quad \ell_1 := \lfloor\tfrac{m}{2}\rfloor \ , \]
\[ \matr{\tilde{L}}_{1\frac{1}{n+1}}^{\infty} = \matr{I}_m - \sum_{k=0}^{\ell_2-1} p_{m-2k-2} \matr{E}_{m-1-k,1+k} \ , \quad \ell_2 := \lfloor\tfrac{m-1}{2}\rfloor \ . \]
When $n+1=2m+1$, we instead have (see Appendix):
\[ \begin{cases} (\tilde{Q}_1^{\infty})_{m+1-k,1+k} + (\tilde{Q}_{1\frac{1}{n+1}}^{\infty})_{m+1+k,2m+1-k} = 0 \ , & \\
(\tilde{Q}_1^{\infty})_{m+2+k,2m+1-k} + (\tilde{Q}_{1\frac{1}{n+1}}^{\infty})_{m-k,1+k} = 0 \ , & \end{cases} \]
and looking ahead to Proposition \ref{StokesCongruency}, it will be convenient to define:
\begin{equation}\label{OddCaseEntries}
p_{m-2k} := (-1)^{m} (\tilde{Q}_1^{\infty})_{m+1-k,1+k} \ , \quad p_{m-2k-1} := (-1)^{m-1}(\tilde{Q}_{1\frac{1}{n+1}}^{\infty})_{m-k,1+k} \ .
\end{equation}
Then:
\[ \tilde{\matr{Q}}_1^{\infty} = 
\left[ \begin{array}{c | c}
\matr{\tilde{L}}_1^{\infty} & \vect{0} \vphantom{\bigg( \bigg)} \\ \hline
\vect{0} & \vphantom{\bigg( \bigg)} \big( \matr{\tilde{L}}_{1\frac{1}{n+1}}^{\infty} \big)^{-\matr{T}} 
\end{array} \right] \ , \quad
\tilde{\matr{Q}}_{1\frac{1}{n+1}}^{\infty} = 
\left[ \begin{array}{c | c}
\matr{\tilde{L}}_{1\frac{1}{n+1}}^{\infty} & \vect{0} \vphantom{\bigg( \bigg)} \\ \hline
\vect{0} & \vphantom{\bigg( \bigg)} \big(\matr{\tilde{L}_1}^{\infty}\big)^{-\matr{T}}
\end{array}\right] \ , \]
where:
\[ \matr{\tilde{L}}_1^{\infty} = \matr{I}_{m+1} + \sum_{k=0}^{\ell_1'-1} p_{m-2k} \matr{E}_{m+1-k,1+k} \ , \quad \ell_1' := \lfloor\tfrac{m+1}{2}\rfloor \ , \]
\[ \matr{\tilde{L}}_{1\frac{1}{n+1}}^{\infty} = \matr{I}_m + \sum_{k=0}^{\ell_2'-1} p_{m-2k-1} \matr{E}_{m-k,1+k} \ , \quad \ell_2' := \lfloor\tfrac{m}{2}\rfloor \ . \]

To facilitate the next few propositions, we introduce the permutation matrix $\Delta := \sum_{k=0}^{m-2} \matr{E}_{m-1-k,1+k}$ and the block-matrix:
\begin{equation}\label{FMatrix}
\matr{F} := \begin{cases} 
\ \left[ \begin{array}{c | c} \matr{L}^{[m]} & \vect{0} \\ \hline \vect{0} \vphantom{\big( \big)} & \matr{U}^{[m]} \end{array} \right] \ , & n+1=2m   \ , \\
  \left[ \begin{array}{c | c} \matr{L}^{[m+1]} & \vect{0} \\ \hline \vect{0} \vphantom {\big( \big)} & \matr{U}^{[m]} \end{array} \right] \ , & n+1=2m+1 \ , \end{cases}
\end{equation}
where $\matr{L}^{[m]}$ and $\matr{U}^{[m]}$ are defined as follows (here, $\ell_1 = \lfloor\tfrac{m}{2}\rfloor$ and $\ell_2 = \lfloor\tfrac{m-1}{2}\rfloor$):
\[ \matr{L}^{[m]} := \matr{I}_m + \sum_{k=0}^{\ell_1-1} \varepsilon p_{m-2k-1} \sum_{j=0}^{k} \matr{E}_{m-2k+j,j+1} + \sum_{k=0}^{\ell_2-1} \varepsilon p_{m-2k-2} \sum_{j=0}^{k} \matr{E}_{m-2k-1+j,j+1} \ , \]
\[ \matr{U}^{[m]} := \left[ \begin{array}{c | c} 
1 & \vect{0} \\ \hline
\vect{0} \vphantom{\Big( \Big)} & \Delta\matr{L}^{[m-1]}\Delta 
\end{array} \right] \ . \]
For example, when $m=4$:
\[ \matr{L}^{[4]} = \begin{bmatrix} 1 & & & \\ \varepsilon p_1 & 1 & & \\ \varepsilon p_2 & \varepsilon p_1 & 1 & \\ \varepsilon p_3 & 0 & 0 & 1 \end{bmatrix} \ , \quad
\matr{U}^{[4]} = \begin{bmatrix} 1 & 0 & 0 & 0 \\ & 1 & 0 & \varepsilon p_2 \\ & & 1 & \varepsilon p_1 \\ & & & 1 \end{bmatrix} \ . \]
When $m=1$, we define $\matr{L}^{[1]} := 1$ and $\matr{U}^{[1]} := 1$, and for the degenerate cases $m \leq 0$, we let $\matr{L}^{[m]}$ be the empty ($0 \times 0$) matrix.  This allows us to state the following lemma:
\begin{lemma}
The following block-matrix identities hold for all $m \geq 1$:
\begin{equation}\label{TriangleSplit}
\matr{U}^{[m]} \matr{L}^{[m] \, \matr{T}} = \begin{bmatrix} 1 & \varepsilon p_1 & \cdots & \varepsilon p_{m-2} & \varepsilon p_{m-1} \\
                                                              &          1      & \ddots &         \ddots      & \varepsilon p_{n-1} \\
                                                              &                 & \ddots &         \ddots      &         \vdots      \\
                                                              &                 &        &            1        &   \varepsilon p_1   \\
                                                              &                 &        &                     &            1        \end{bmatrix} \ .
\end{equation}
\begin{equation}\label{UniversalSequence1}
\matr{F}\matr{\tilde{Q}}_1^{\infty} = \left\{ \begin{array}{cl}
\left[ \begin{array}{c | c}
\matr{L}^{[m-1]} & \vect{0} \\ \hline
\vect{0} \vphantom{\Big( \Big)} & \matr{U}^{[m+1]}
\end{array} \right] \ , & n+1=2m \ , \\
\left[ \begin{array}{c | c}
\matr{L}^{[m]} & \vect{0} \\ \hline
\vect{0} & \vphantom{\Big( \Big)} \matr{U}^{[m+1]} 
\end{array} \right] \ , & n+1=2m+1 \ ,
\end{array}\right.
\end{equation}
\begin{equation}\label{UniversalSequence2}
\matr{F}\matr{\tilde{Q}}_1^{\infty}\matr{\tilde{Q}}_{1\frac{1}{n+1}}^{\infty} = \left\{ \begin{array}{cl} 
\left[ \begin{array}{c | c} 
\matr{L}^{[m-2]} & \vect{0} \\ \hline 
\vect{0} & \vphantom{\Big( \Big)} \matr{U}^{[m+2]} 
\end{array} \right] \ , & n+1=2m \ , \\
\left[ \begin{array}{c | c}
\matr{L}^{[m-1]} & \vect{0} \\ \hline
\vect{0} & \vphantom{\Big( \Big)} \matr{U}^{[m+2]} 
\end{array} \right] \ , & n+1=2m+1 \ .
\end{array}\right.
\end{equation}
\end{lemma}

\prf These follow directly from (\ref{FMatrix}), (\ref{EvenCaseEntries}), and (\ref{OddCaseEntries}). \quad $\blacksquare$ \\

To make the connection with Section 2, we now define $\matr{R}$ as in Equation (\ref{RMatrix}) using the entries $p_1, \dots, p_m$ defined via (\ref{EvenCaseEntries}) or (\ref{OddCaseEntries}), and by defining $p_{n+1-k} := \varepsilon p_k$. Then:
\begin{prop}\label{RSimilar}
$\matr{R} = \matr{F}\mathcal{R}\matr{F}^{-1}$.  Hence:
\[ p(x) := \det(x\matr{I}_{n+1} - \mathcal{R}) = x^{n+1} + \sum_{k=1}^{n} p_k x^k + \varepsilon \ , \quad p_{n+1-k} = \varepsilon p_k \ . \]
\end{prop}

\prf Using (\ref{UniversalSequence2}), we obtain $\matr{R F}\matr{R}_{\varepsilon}^{-1} = \matr{F}\tilde{\matr{Q}}_1^{\infty}\tilde{\matr{Q}}_{1\frac{1}{n+1}}^{\infty}$. \quad $\blacksquare$

\begin{prop}\label{StokesCongruency}
Let $\matr{S} := \matr{F}\mathcal{S}\matr{F^T}$. Then:
\[ \matr{S} = \begin{bmatrix} 1 & \varepsilon p_1 & \cdots & \varepsilon p_{n-1} &   \varepsilon p_n   \\
                                &          1      & \ddots &         \ddots      & \varepsilon p_{n-1} \\
                                &                 & \ddots &         \ddots      &         \vdots      \\
                                &                 &        &            1        &   \varepsilon p_1   \\
                                &                 &        &                     &            1        \end{bmatrix} \ . \]
\end{prop}

\prf Letting $\matr{J}_p^{-1} := \matr{F}\mathcal{J}^{-1}\matr{F^T}$, we find, by definition (\ref{RealStokesMatrix}) of $\mathcal{S}$ and Lemma \ref{RSimilar}, that $\matr{S} = \matr{R}^m\matr{J}_p^{-1}$.  Hence, by inspection when $n+1=2m$, and by (\ref{UniversalSequence1}) when $n+1=2m+1$:
\[ \matr{J}_p^{-1} = \left\{ \begin{array}{cl}
\left[\begin{array}{c | c}
 \vect{0} & -\varepsilon \matr{L}^{[m]}\matr{U}^{[m] \, \matr{T}} \\ \hline
\vphantom{\Big( \Big)} \matr{U}^{[m]}\matr{L}^{[m] \, \matr{T}} & \vect{0}
\end{array}\right] \ , & n+1=2m \ , \\
\left[\begin{array}{c | c}
\vect{0} & -\varepsilon \matr{L}^{[m]}\matr{U}^{[m] \, \matr{T}} \\ \hline
\vphantom{\Big( \Big)} \matr{U}^{[m+1]}\matr{L}^{[m+1] \, \matr{T}} & \vect{0} 
\end{array}\right] \ , & n+1=2m+1 \ .
\end{array} \right. \]
Since $\matr{R}\vect{e}_k=\vect{e}_{k-1}$ for all $1 \leq k \leq n$ (cf. proof of Proposition 2.1), we see that $\matr{R}^m\matr{J}_p^{-1}\vect{e}_1 = \vect{e}_1$, and:
\[ \matr{R}^m\matr{J}_p^{-1}\vect{e}_k = p_{k-1}\vect{e}_1 + \cdots + p_1\vect{e}_{k-1} + \vect{e}_k \ , \quad \left\{\begin{array}{ll} 
2 \leq k \leq m \ , & n+1=2m \\
2 \leq k \leq m+1 \ , & n+1=2m+1 \ . \end{array}\right. \]
This determines the first $m$ ($m+1$) columns of $\matr{S}$; we now determine the remaining $m$ columns by applying $\matr{S}$ to the flag:
\[ \mathcal{F} \ : \ \vect{0} \subset \mathcal{F}_0 \subset \mathcal{F}_1 \subset \cdots \subset \mathcal{F}_{n} \cong \R^{n+1} \ , \quad \mathcal{F}_k := \, < \vect{e}_{n+1}, \dots, \vect{e}_{n+1-k} >_\R \ . \]
From (\ref{TriangleSplit}), we observe that $-\varepsilon\matr{J}_p^{-1}\vect{e}_{n+1} = \vect{e}_m$, and that:
\[ -\varepsilon\matr{J}_p^{-1}\vect{e}_{n+1-k} - \vect{e}_{m-k} \in \, <\vect{e}_m, \dots, \vect{e}_{m-k+1}>_\R \ , \quad \forall \ 1 \leq k \leq m-1 \ . \]
As a result, for all $0 \leq k \leq m-1$:
\[ -\varepsilon\matr{J}_p^{-1}\mathcal{F}_k = <\vect{e}_m, \dots, \vect{e}_{m-k}>_\R = \left\{ \begin{array}{cl}
\matr{R}^m\mathcal{F}_k \ , & n+1=2m \ , \\
\matr{R}^{m+1}\mathcal{F}_k \ , & n+1=2m+1 \ , \end{array}\right. \]
and hence:
\[ \matr{R}^m\matr{J}_p^{-1}\mathcal{F}_k = -\varepsilon\matr{R}^{n+1}\mathcal{F}_k = \matr{U}\mathcal{F}_k \ , \]
where $\matr{U}$ was found in the proof of Proposition \ref{NthRoot}. Therefore, the last $m$ columns of $\matr{S}$ are the last $m$ columns of $\matr{U}$, and this concludes the proof. \quad $\blacksquare$

\begin{thm}
$\mathcal{S} + \mathcal{S}^\matr{T}$ has the same signature as $\matr{diag} \, ( \varepsilon p(\pi_0 ), \dots, \varepsilon p(\pi_{n}) )$.
\end{thm}

\prf By Proposition \ref{StokesCongruency}, $\matr{S+S^T}$ and $\mathcal{S} + \mathcal{S}^\matr{T}$ are congruent via $\matr{F^T}$.  Since real symmetric matrices are diagonalizable, congruence implies that they have equal rank and signature, and thus, the theorem follows by Propositions \ref{RSimilar} and \ref{SignatureProp}. \quad $\blacksquare$

To conclude, we state what this means in terms of solutions to the $tt^*$-Toda equation (\ref{tt*-TodaEqtn}).  It was shown in \cite{GIL1, GIL2, GL1, MoXX} that solutions $w_i:\C^* \rightarrow \R$ are in one-to-one correspondence with real numbers $\gamma_i$ satisfying $\gamma_i - \gamma_{i-1} \geq -2$ for all $i$, where $2w_i(z) \sim \gamma_i \log|z|$ as $|z| \rightarrow 0$.  When $n+1=2m$, the corresponding eigenvalues of $\matr{R}$ are $exp\big( \pm\tfrac{i\pi}{n+1}(\gamma_j + 2j + 1) \big)$, $0 \leq j \leq m-1$, with
\[ 0 \leq \tfrac{\pi}{n+1}(\gamma_0 + 1) \leq \tfrac{\pi}{n+1}(\gamma_1 + 3) \leq \cdots \leq \tfrac{\pi}{n+1}(\gamma_{m-1} + 2m - 1) \leq \pi \ . \]
The condition $\mathcal{S}+\mathcal{S}^\matr{T} > 0$ means that these points must interlace with the $(n+1)^\matr{th}$ roots of unity:
\[ 0 < \gamma_0 + 1 < 2 < \gamma_1 + 3 < 4 < \cdots < 2m-2 < \gamma_{m-1} + 2m - 1 < 2m \ , \]
and this means that $|\gamma_j| < 1$ for all $j=0,\dots,m-1$.

\section{Appendix: The Stokes factor matrices for the case $n+1=2m+1$}

We present the details of the $n+1=2m+1$ case, some of which require separate calculation from the $n+1=2m$ case.  We first recall, from Section 3 and (\ref{StokesRays}), that the potentially non-zero entries of $\matr{Q}_1^{\infty}$ and $\matr{Q}_{1\frac{1}{n+1}}^{\infty}$ are:
\[ (Q_1^{\infty})_{m+1-k,1+k} \ , \ (Q_1^{\infty})_{m+2+k,2m+1-k} \ , \ (Q_{1\frac{1}{n+1}}^{\infty})_{m-k,1+k} \ , \ \textrm{and} \ (Q_{1\frac{1}{n+1}}^{\infty})_{m+1+k,2m+1-k} \ , \]
for $k \geq 0$ stopping before the diagonal.  Thus, $\matr{Q}_1^{\infty}$ and $\matr{Q}_{1\frac{1}{n+1}}^{\infty}$ have the following block structure:
\[ \matr{Q}_1^{\infty} = \left[ \begin{array}{c | c}
\matr{L}_1 & \vect{0} \\ \hline
\vect{0} & \matr{U}_1 \end{array} \right] \ , \quad 
\matr{Q}_{1\frac{1}{n+1}}^{\infty} = \left[ \begin{array}{c | c}
\matr{L}_{1\frac{1}{n+1}} & \vect{0} \\ \hline
\vect{0} & \matr{U}_{1\frac{1}{n+1}} \end{array} \right] \ , \]
where ($\ell_1 := \lfloor\tfrac{m+1}{2}\rfloor$, $\ell_2 := \lfloor\tfrac{m}{2}\rfloor$):
\[ \matr{L}_1 := \matr{I}_{m+1} + \sum_{k=0}^{\ell_1-1} \alpha_{m-2k} \matr{E}_{m+1-k,1+k} \ , \quad
\matr{L}_{1\frac{1}{n+1}} := \matr{I}_m + \sum_{k=0}^{\ell_2-1} \alpha_{m-2k-1} \matr{E}_{m-k,1+k} \ , \]
and where the upper-triangular matrices $\matr{U}_1$ and $\matr{U}_{1\frac{1}{n+1}}$ will be expressed in terms of $\matr{L}_{1\frac{1}{n+1}}$ and $\matr{L}_1$, respectively, using the following symmetries.  Let $\omega := e^{ \big( \frac{2\pi i}{n+1} \big) }$, $\matr{d}_{n+1} := \matr{diag} \, (1, \omega, \dots, \omega^n)$, $\Pi$ be as before, and $\Delta := \sum_{k=0}^{n-1} \matr{E}_{n-k,1+k}$.  Recall that for all $k \in \tfrac{1}{n+1}\Z$:
\begin{enumerate}
\item $\Z_{n+1}$-symmetry: $\matr{Q}_{k+\frac{2}{n+1}}^{\infty} = \Pi\matr{Q}_k^{\infty}\Pi^{-1}$.
\item $\llq$Anti-symmetry": $\matr{Q}_{k+1}^{\infty} = \matr{d}_{n+1}^{-1}\big( \matr{Q}_k^{\infty} \big)^{-\matr{T}} \matr{d}_{n+1}$.
\item Reality (at $\zeta=\infty$): $\matr{Q}_k^{\infty} = (\Delta\Pi)^{-1} \big( \overline{\matr{Q}_{\frac{2n+1}{n+1}-k}^{\infty}} \, \big)^{-1}  \Delta\Pi$.
\end{enumerate}
Then by $\Z_{n+1}$-symmetry, and anti-symmetry for $k=1$:
\[ \Pi^m \matr{Q}_{1\frac{1}{n+1}}^{\infty}\Pi^{-m} = \matr{Q}_2^{\infty} = \matr{d}_{n+1}^{-1}\big( \matr{Q}_1^{\infty} \big)^{-\matr{T}}\matr{d}_{n+1} \ , \]
and hence:
\[ \matr{Q}_{1\frac{1}{n+1}}^{\infty} = \big( \matr{d}_{n+1} \Pi^m \big)^{-1} \big( \matr{Q}_1^{\infty} \big)^{-\matr{T}} \big( \matr{d}_{n+1} \Pi^m \big) \ . \]
Moreover, by the reality condition for $k=1$, and $\Z_{n+1}$-symmetry for $k=\tfrac{n}{n+1}$:
\[ \matr{Q}_1^{\infty} = (\Delta\Pi)^{-1} \big( \overline{ \matr{Q}_{\frac{n}{n+1}}^{\infty} } \big)^{-1} (\Delta\Pi) = (\Delta\Pi)^{-1} \Pi^{-1} \big( \overline{ \matr{Q}_{1\frac{1}{n+1}}^{\infty} } \big)^{-1} \Pi (\Delta\Pi) = \Delta \big( \overline{ \matr{Q}_{1\frac{1}{n+1}}^{\infty} } \big)^{-1} \Delta \ . \]
The block structure of $\matr{Q}_{1\frac{1}{n+1}}^{\infty}$ then implies that:
\[ \matr{Q}_1^{\infty} = \left[ \begin{array}{c | c}
\overline{\matr{U}_{1\frac{1}{n+1}}}^{\, -\matr{T}} & \vect{0} \\ \hline
\vect{0} \vphantom{\bigg(} & \overline{\matr{L}_{1\frac{1}{n+1}}}^{\, -\matr{T}} \end{array} \right] \ , \]
and hence:
\[ \matr{U}_1 = \overline{ \matr{L}_{1\frac{1}{n+1}} }^{\, -\matr{T}} \ , \quad \matr{U}_{1\frac{1}{n+1}} = \overline{ \matr{L}_1 }^{\, -\matr{T}} \ . \]
In addition, using the above identities, we deduce that:
\[ \big( \Pi^m\Delta \big) \matr{Q}_1^{\infty} \big( \Pi^m\Delta \big)^{-1} = \matr{d}_{n+1} \big( \matr{Q}_1^{\infty} \big)^* \matr{d}_{n+1}^{-1} \]
\[ \Leftrightarrow \quad \left\{ \begin{array}{cl}
\matr{L_1^T} & = \matr{diag} \, (1,\omega,\dots,\omega^m) \, \matr{L}_1^* \, \matr{diag} \, (1,\omega,\dots,\omega^m)^{-1} \ , \\
\matr{U_1^T} & = \matr{diag} \, (1,\omega,\dots,\omega^{m-1}) \, \matr{U}_1^* \, \matr{diag} \, (1,\omega,\dots,\omega^{m-1})^{-1} \ . \end{array}\right. \]
Using the entry-wise expression of $\matr{L}_1$ and $\matr{U}_1$, this implies that for all $k \geq 0$:
\[ \overline{\alpha}_{m-2k} = \alpha_{m-2k}\omega^{m-2k} \ , \quad \overline{\alpha}_{m-2k-1} = \alpha_{m-2k-1}\omega^{m-2k-1} \ . \]
Consequently, we see that:
\[ \alpha_{m-2k} = |\alpha_{m-2k}| (\omega^{-\frac{1}{2}})^{m-2k} \ , \quad \alpha_{m-2k-1} = |\alpha_{m-2k-1}|(\omega^{-\frac{1}{2}})^{m-2k-1} \ . \]

We now wish to determine the exponent $r$ for which the following matrices are real-valued:
\[ \matr{d}_{n+1}^r \matr{Q}_1^{\infty} \matr{d}_{n+1}^{-r} \ , \quad
\matr{d}_{n+1}^r \matr{Q}_{1\frac{1}{n+1}}^{\infty} \matr{d}_{n+1}^{-r} \ , \]
such that $\matr{d}_{n+1}^r \Pi = \Pi \matr{d}_{n+1}^r$.  Evidently, the non-zero entries of these matrices are:
\[ |\alpha_{m-2k}| \omega^{(m-2k)(r-\frac{1}{2})} \ , \quad |\alpha_{m-2k-1}| \omega^{(m-2k-1)(r-\frac{1}{2})} \ , \]
and these are real \emph{iff} for all $k$:
\[ (m-2k)(r-\tfrac{1}{2}) \equiv 0 \mod \tfrac{n+1}{2} \ , \quad (m-2k-1)(r-\tfrac{1}{2}) \equiv 0 \mod \tfrac{n+1}{2} \ . \]
Of the two solutions $r=\tfrac{1}{2}$ and $r=m+1$, only $r=m+1$ satisfies $\matr{d}_{n+1}^r \Pi = \Pi \matr{d}_{n+1}^r$.  As such, we define:
\[ \matr{\tilde{Q}}_1^{\infty} := \matr{d}_{n+1}^{m+1} \matr{Q}_1^{\infty} \matr{d}_{n+1}^{-m-1} \ , \quad
\matr{\tilde{Q}}_{1\frac{1}{n+1}}^{\infty} := \matr{d}_{n+1}^{m+1} \matr{Q}_{1\frac{1}{n+1}}^{\infty} \matr{d}_{n+1}^{-m-1} \ , \]
which implies that:
\begin{align*}
\matr{\tilde{L}}_1 & := \matr{diag} \, (1,\omega,\dots,\omega^m)^{m+1} \ \matr{L}_1 \ \matr{diag} \, (1,\omega,\dots,\omega^m)^{-m-1} \\
 & = \matr{I}_{m+1} + \sum_{k=0}^{\ell_1-1} (-1)^m |\alpha_{m-2k}| \ \matr{E}_{m+1-k,1+k} \ , 
\end{align*}
\begin{align*}
\matr{\tilde{L}}_{1\frac{1}{n+1}} & := \matr{diag} \, (1, \omega,\dots,\omega^{m-1})^{m+1} \ \matr{L}_{1\frac{1}{n+1}} \ \matr{diag} \, (1,\omega,\dots,\omega^{m-1})^{-m-1} \\
 & = \matr{I}_m + \sum_{k=0}^{\ell_2-1} (-1)^{m-1} |\alpha_{m-2k-1}| \ \matr{E}_{m-k,1+k} \ . \quad \spadesuit
\end{align*}

\end{document}